%% file: main.tex
\documentclass[journal]{IEEEtran}

\usepackage{graphicx} 
\usepackage{amsmath,amsfonts,amssymb}
\usepackage{graphicx}
\usepackage{xcolor}
\usepackage{multicol}
\usepackage{flushend}
\usepackage{algorithm}
\usepackage{algorithmic}
\usepackage{tikz}
\usepackage{authblk}
\usetikzlibrary{arrows.meta}

\usepackage[switch]{lineno}

\input{math-definitions.tex} 

\input{acronyms}

\title{Real-Time 3D Magnetic Field Camera for a Spherical Volume}

\author[1,2,*]{Fynn Foerger}
\author[1,2]{Marija Boberg}
\author[1,2]{Niklas Hackelberg}
\author[4,5,6]{Philip Heinisch}
\author[4,5,6]{Katharina Ostaszewski}
\author[1,2]{Jonas Faltinath}
\author[1,2]{Florian Thieben}
\author[1,2]{Fabian Mohn}
\author[1,2]{Paul Jür\ss}
\author[1,2]{Martin Möddel}
\author[1,2,3]{Tobias Knopp}

\affil[1]{Section for Biomedical Imaging, University Medical Center Hamburg-Eppendorf, Hamburg, Germany}
\affil[2]{Institute for Biomedical Imaging, Hamburg University of Technology, Hamburg, Germany}
\affil[3]{Fraunhofer Research Institution for Individualized and Cell-based Medical Engineering IMTE, Lübeck, Germany}
\affil[4]{Institut für angewandte numerische Wissenschaft, Braunschweig, Germany}
\affil[5]{Institut für Geophysik und extraterrestrische Physik, Technische Universität Braunschweig, Braunschweig, Germany}
\affil[6]{PhySens GmbH, Braunschweig, Germany}

\affil[*]{Corresponding author, \texttt{fynn.foerger@tuhh.de}}

\begin{document}

\maketitle

\begin{abstract}
    Accurate and efficient volumetric magnetic field measurements are essential for a wide range of applications. Conventional methods are often limited in terms of measurement speed and applicability, or suffer from scaling problems at larger volumes.
    This work presents the development of a magnetometer array designed to measure magnetic fields within a spherical volume at a frame rate of 10\,Hz. The array consists of 3D Hall magnetometers positioned according to a spherical $\bm t$-design, allowing simultaneous magnetic field data acquisition from the surface of the sphere. The approach enables the efficient representation of all three components of the magnetic field inside the sphere using a sixth-degree polynomial, significantly reducing measurement time compared to sequential methods. This work details the design, calibration, and measurement methods of the array. To evaluate its performance, we compare it to a sequential single-sensor measurement by examining a magnetic gradient field. The obtained measurement uncertainties of approx. 1\% show the applicability for a variety of applications.
\end{abstract}

\section{Introduction}
\begin{figure*}[]
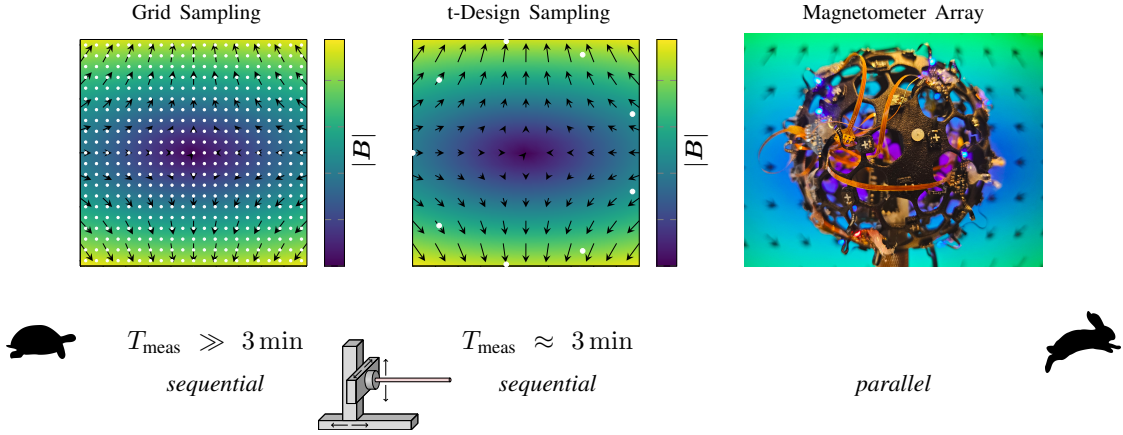

    \centering
    \include{tikz/1-intro_motivation}
    \caption{Schematic representation of different magnetic field measurement methods. Depending on the measurement method, different measurement times $T_{meas}$ and field accuracies are achieved. The specified times refer to empirical values for measurements of magnetic fields in a spherical measurement volume with a radius of \SI{4.5}{cm}. The simplest method, shown on the left, involves moving a single 3D Hall magnetometer sequentially to grid points within the measurement volume using a robot, with the white dots indicating the measurement positions. This procedure can be very time-consuming, especially for 3D measurements and is limited by the speed of the mechanical movement of the robot. In the next step, prior knowledge of the underlying magnetic field equations is utilized to measure only on a spherical surface, allowing the field inside the sphere to be determined, as illustrated in the middle image. Typically, this method requires several minutes for sequential measurements. In this work, we combine this method with a magnetometer array, illustrated on the right, which enables parallel field readout, resulting in volumetric magnetic field measurements at a rate of several Hertz.}
    \label{fig:Intro}
\end{figure*}

Magnetic fields play a crucial role in various areas, ranging from consumer products over industrial processes to scientific research and medical diagnostics. Their inherent ability to penetrate matter and interact with it in various ways enables a wide range of applications, including the transfer of forces and torques.
The spatial distribution of magnetic fields is of high concern in most of these applications, as it directly influences the functionality and accuracy of the processes involved. One important area of magnetic fields in healthcare is tomographic imaging, which enabled the development of \ac{MRI}~\cite{lauterbur_image_1989} and \ac{MPI}~\cite{gleich_tomographic_2005}. Both imaging modalities use the superposition of different kinds of magnetic fields for signal generation and encoding. A detailed understanding of the magnetic field configuration is essential, as it directly underpins high-resolution imaging and precise diagnostics~\cite{mukhatov_comprehensive_2023,dietrich_field_2016,knopp_magnetic_2017}. As a result, precise magnetic field measurements are frequently conducted and are essential for accurate imaging.

Often, magnetic field measurements in a volume relied on a single 3D sensor positioned sequentially at grid points~\cite{Boberg2025}. Although this approach is reliable, it is also time-consuming requiring precise and repeated repositioning of the sensor~\cite{akai_3d_2017}. 
To address these challenges, previous research has explored the use of multiple sensors to achieve higher field capture frame rates~\cite{SCHLAGETER200137,10098773,Nicolas2024fpga}. In addition, taking advantage of the fact that in regions of zero current density the field components satisfy the Laplace equation, each component can be expressed by an expansion of spherical harmonics. Using the field boundary values at the surface of the examined volume, the expansion coefficients can be completely determined resulting in a polynomial for the field inside the sphere~\cite{10.1063/1.4872244,ECCLES1993135,bringout_robust,barmet_spatiotemporal_2008,suksmono_magnetic_2021,dietrich_field_2016,Metrolab}. For a spherical volume, by incorporating an efficient quadrature on the surface with points at spherical $t$-design positions~\cite{Beentjes2016QUADRATUREOA,Boberg2025}, fields can be calculated with small uncertainty using only a minimal number of measurement points \cite{Boberg2025}. 

In this work, we aim to extend the approach proposed by Boberg et al.~\cite{Boberg2025} by combining it with a spherical magnetometer array. By simultaneously reading sensor data at the $t$-design positions, we significantly reduce the measurement time compared to the sequential single-sensor approach. With $86$ 3D magnetometers, the result is a polynomial representation of the magnetic field with a degree of $6$ that can be evaluated at any point within the sphere with a diameter of \SI{9}{\cm}.

The choice of magnetometer technology depends largely on the application, with key factors including the required field strength and the fields rate of change~\cite{9296742}. In addition, the size of the available sensor elements and the crosstalk between different sensors must be taken into account for the intended positioning. Various vector magnetic field sensors can measure in the desired frequency range from DC to approximately \SI{100}{\Hz}. These include anisotropic magneto-resistance~\cite{grosz_high_2017}, giant magneto-resistance~\cite{grosz_high_2017}, tunneling magneto-resistance~\cite{ripka_modern_2007}, and giant magneto-impedance~\cite{PHAN2008323} sensors, as well as fluxgate~\cite{grosz_high_2017}  and Hall-sensors~\cite{fraden_handbook_2004}. For AC-only measurements, induction coils can also be used, operating in the sub-1 Hz range~\cite{9296742}. 

Each of these sensor types differs in terms of accuracy, dynamic range, and noise characteristics~\cite{9296742}. Since our magnetometer array is primarily intended for MPI magnetic field sequence planning, it must accommodate field strengths of several hundred \SI{}{\milli\tesla}~\cite{DesignAndOpt}. Hall sensors in particular are well suited for this field range, while remaining cost-effective~\cite{9296742}.

By using Hall magnetometers, our system achieves fast measurements of all three components of the magnetic field, providing an improvement in both speed and functionality while maintaining low hardware complexity compared to previous work. For example, the field camera system developed by Dietrich et al.~\cite{dietrich_field_2016} uses nuclear magnetic resonance (NMR) probes in a spherical configuration to measure the magnitude of the field. However, the NMR sensor requires a magnetic offset field and complex high frequency electronics for signal generation and acquisition, and does not measure the full field vector.
Another commercially available magnetic field measurement system uses NMR probes arranged in a semicircular geometry and requires multiple measurements at different angles to reconstruct the magnetic field~\cite{Metrolab}. As a result, detection times are far from the sub-second range, limiting its applicability.

The spherical magnetometer array opens up new possibilities for real-time applications and facilitates scaling to larger volumes or higher number of sensors due to its reduced hardware complexity. The concept of field measurement is outlined in \cref{fig:Intro}, emphasizing the time savings compared to the single sensor measurement approach.

\section{Theory}

This section outlines the theoretical basis and mathematical framework of the spherical magnetometer array using spherical harmonics to measure the magnetic field across an entire volume in real time. Prior knowledge of the underlying physical relationships is used to determine a field in a volume with a minimum number of field probes. A more detailed analysis can be found in the work of Boberg et al.~\cite{Boberg2025}, where the complete derivation is included.

The fundamental equations governing electromagnetic phenomena are unified in Maxwell's equations. From these, in the quasi-static limit and in the absence of current densities, the magnetic flux density $\bm{B}: \mathbb{S}_R  \rightarrow \IR^3$ in the spherical volume under consideration $\mathbb{S}_R := \left\{ \bm a \in \mathbb{R}^3 : \|\bm a \|_2 \leq R \right\}$, with radius $ R \in \mathbb{R}_+$, is described by
\begin{align*}
    \nabla \times \bm B \left(\bm r\right) &= \bm 0 \\ 
    \nabla \cdot \bm B \left(\bm r\right) &= \bm 0.
\end{align*}
It follows from these two equations that $\bm B$ satisfies Laplace's equation
\begin{align*}
    \Delta B_i(\bm{r}) = 0, \quad i \in \{x, y, z\}
\end{align*}
for each component. This fundamental property allows for the expansion of the magnetic flux density in terms of spherical harmonics
\begin{align}
    B_i(\bm{r}) = \sum_{l=0}^{\infty} \sum_{m=-l}^l \gamma_{l,m}^i Z_l^m(\bm{r}) \quad \forall \bm{r} \in \mathbb{S}_R 
    \label{eq:expansion}
\end{align}
where $\gamma_{l,m}^i \in \IR$ are the expansion coefficients and $Z_l^m(\bm{r})$ the solid spherical harmonic functions with
\begin{align*}
    & Z^m_l : \IR^3 \rightarrow \IR, \\ & (r,\theta,\phi) \mapsto K^{|m|}_l r^l P^{|m|}_l\left(\cos\left(\theta\right)\right)\begin{cases} \sqrt{2}\cos\left(m\phi\right) & m > 0 \\ \sqrt{2}\sin\left(|m|\phi\right) & m < 0\\ 1 & m=0\end{cases} .
\end{align*}
Here, we have $K^m_l = \sqrt{\frac{(l-m)!}{(l+m)!}}$, and the associated Legendre polynomials $P^m_l$.
The coefficients $\gamma_{l,m}^i$ can be calculated by applying the Dirichlet boundary condition
\begin{align}
    \gamma_{l,m}^i &= \frac{2l+1}{R^l 4\pi} \int_{\mathbb{S}_R} B_i(\bm{r}) Z_l^m\left(\frac{\bm{r}}{R}\right) \, d\bm{r} .
    \label{eq:coeffsInt}
\end{align}

The integral in equation \eqref{eq:coeffsInt} is solved by using a quadrature that features spherical $t$-design positions $\{\bm{r}_1, \dots, \bm r_N \} \subseteq \partial\mathbb{S}_R$ as sampling points of the integral:
\begin{align}
    \gamma_{l,m}^i = \frac{2l+1}{R^l N} \sum_{k=1}^N B_i(\bm{r}_k) Z_l^m\left(\frac{\bm{r}_k}{R}\right)
    \label{eq:coeffsSum}
\end{align}
In addition to the property that the quadrature weights for a $t$-design are the same for all points, it also holds that the expansion in \ref{eq:expansion} with the approximation using the $t$-design quadrature in \eqref{eq:coeffsSum} is exact if the integrand has a polynomial degree of at most $t$. The consequence is that the field can be expressed exactly inside the sphere if the polynomial degree of $\bm B$ is $\lfloor \frac{t}{2} \rfloor$ or less.

Besides the spherical $t$-design, there are several other types of quadrature that can be considered, most notably the class of Gaussian quadratures~\cite{weber_behandlung_2016}. For example, the combination of the trapezoidal rule and the Gauss-Legendre quadrature can provide an exact approximation for polynomials of the same degree as the $t$-design, but with a higher number of nodes \cite{weber_behandlung_2016,Boberg2025}. In addition, such Gaussian quadratures are limited by the physical size of the sensors, as nodes accumulate near the poles. Another advantage is that $t$-designs offer greater robustness to variations in sensor placement and measurement errors~\cite{Scheffler2025Efficient}.

\section{Methods}
The theory described in the last section is now applied to a measuring device for determining magnetic fields in a spherical volume. For this purpose, a magnetometer array consisting of Hall magnetometers is considered in which the field is measured simultaneously at positions defined by a spherical $t$-design.

\subsection{Magnetometer Array}
\begin{figure}
\includegraphics[width=0.24\textwidth]{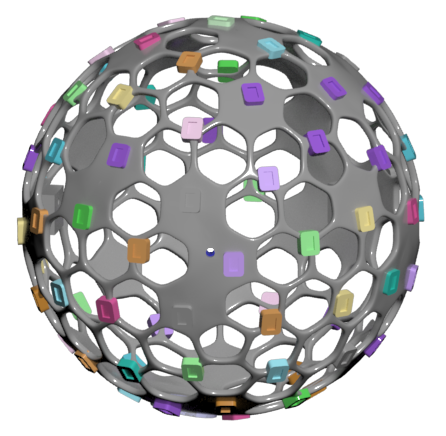}
\includegraphics[width=0.24\textwidth]{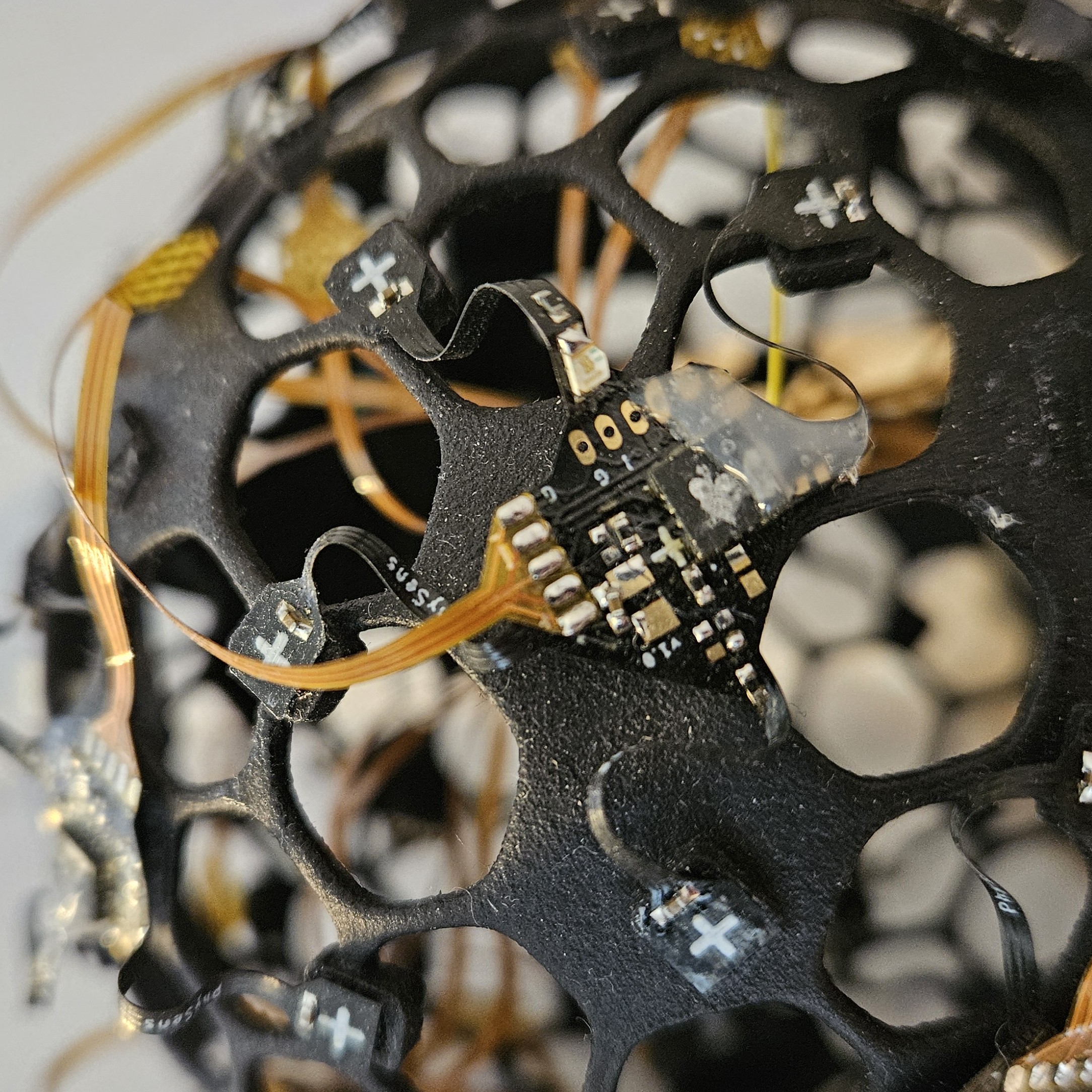}
    \centering
    \caption{Left: 3D rendering of the CAD model of the spherical magnetometer array. The colored rectangular structures provide mechanical support for the individual magnetometers and ensure correct alignment. Hexagonal cutouts allow for easy routing of the flexible printed circuit boards used to connect the magnetometers while still providing the necessary mechanical stiffness. Right: Close up of a flexible sensor printed circuit board. }
    \label{fig:sph_render}
\end{figure}

The sensor array consists of $86$ independent Hall magnetometers (Texas Instruments TMAG5273) arranged on the surface of a 3D printed sphere as shown in Fig. \ref{fig:sph_render} with a radius of $\SI{45}{\mm}$ made of polyamide. To conform to the spherical geometry, the individual sensors are mounted on flexible printed circuit boards with a polyimide film substrate that also serves as electrical connection between the sensors and the central microcontroller. The positions of the individual sensors were chosen based on a $t$-design with $t=12$ and the magnetometer local $y$-axis aligned longitudinally and the local $z$-axis aligned to the position dependent surface normal. 
To reduce the possibility of interference, the power supply, data processing and communications are located on a separate circuit board in the center of the sphere with a shielded non-magnetic cable used for the USB connection. Multiple temperature sensors are available to allow for temperature compensation of the magnetometers. All upstream communications are handled via a 12 Mbps Full-Speed USB interface to further reduce high-frequency electromagnetic interference. 
The magnetometers have a total range of \SI{\pm266}{\milli\tesla} with a resolution of \SI{8.12}{\micro\tesla}. They are sampled simultaneously and internally averaged to achieve an effective data rate of \SI{10}{Hz}. This reduces sensor noise while also conserving transfer bandwidth and computational requirements on the host system. According to the sensors data sheet, sampling rates up to \SI{10}{\kilo\Hz} can be achieved~\cite{TMAG5273}.

\subsection{Calibration Method}
Hall magnetometer, by their nature, are sensitive to various external and internal factors that can affect their readings, hence a careful calibration is important, and the existing literature contains a variety of methods for achieving this~\cite{9296742}. For the Hall magnetometer array, one additional challenge is the uncertainty regarding the exact position and orientation of each sensor within the array. Without a careful calibration, the collected data can be misinterpreted, leading to significant errors of the field representation. First, we describe how to transfer the measured values of the individual sensors into a shared coordinate system. Then we will discuss the calculation of the exact sensor position on the sphere, which we need in order to calculate the coefficients in \eqref{eq:coeffsSum}.
\subsubsection{Local Sensor Calibration}
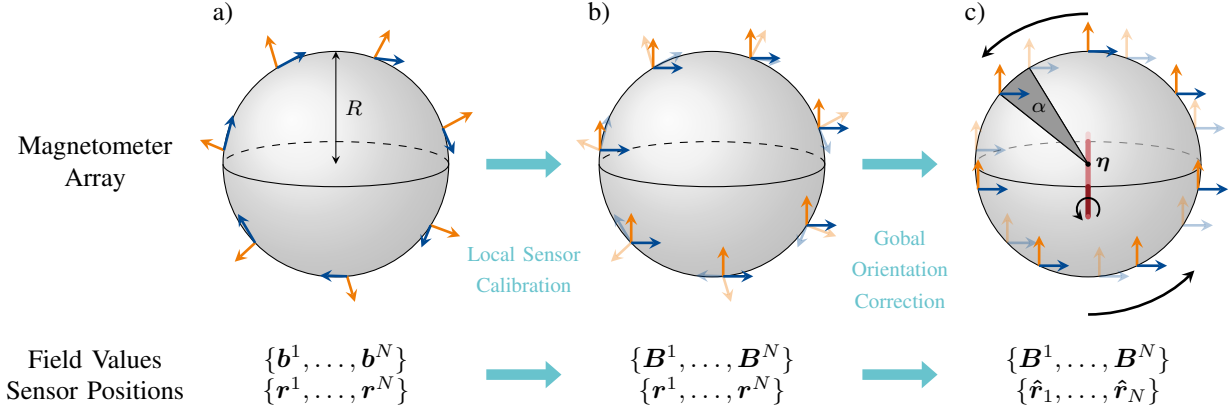
\begin{figure*}
\centering
    \input{tikz/2-theory_calibration}
    \caption{Calibration procedure for the magnetometer array. As shown in a), each Hall magnetometer at $\{\bm r^1,\dots, \bm r^N\}$ outputs its value in a local coordinate system $\{\bm b^1, \dots, \bm b^N\}$ that may not be orthogonal. From a) to b), all local coordinates systems must be transformed into a global coordinate system obtaining $\{ \bm B^1,\dots, \bm B^N\}$ to solve the surface integral for calculation of the expansion coefficients. In a second step from b) to c), the corrected $t$-design positions $\{ \bm{\tilde{r}}_1, \dots, \bm{\tilde{r}}_N \}$ in the coordinate system of the calibration fields are found by applying a global rotation around the axis $\bm \eta$ by the angle $\alpha$}
    \label{fig:calib}
\end{figure*}

As it can be seen in \cref{fig:calib}~a), each sensor in the array possesses its own local coordinate system in which it outputs its field values $\bm b^n$, where the index 
$n$ denotes the sensor number. In the first part of the calibration, the values $\bm b^n$ from the local coordinate systems have to be transformed into a global coordinate system, resulting in values denoted by $\bm B^n$ illustrated in \cref{fig:calib} b). Since the sensor operates within the specified linear range, a linear calibration model can be applied, expressed as
\begin{align*}
\bm B^n = \bm R^n \bm b^n + \bm O^{n} .
\end{align*}
Here, $\bm R^n \in \IR^{3\times 3}$ denotes the individual matrix of the $n$-th sensor for scaling, skew, and rotation and $\bm O^n \in \IR^3$ represents the offset correction. 

To determine both, the rotation matrix and the offset correction, linearly independent homogeneous magnetic fields with known strength and direction in the global coordinate system are applied to the entire device. To obtain a unique solution in this calibration model, $J \geq 4$ different calibration fields spanning $\IR^3$ must be set, denoted by $\Tilde{\bm B}_j \in \IR^3$, where $j$ is the index of the respective field set. $\bm b_j^n$ are the corresponding measured values of the $n$-th sensor when the $j$-th field is applied. To determine $\bm R^n$ and $\bm O^n$, the optimization problem 
\begin{align}
    \argmin_{\bm R^n , \bm O^n} \sum_{j=1}^{J}
    \left\|
    \Tilde{\bm B}_j -\begin{pmatrix} \bm R^n & \bm O^n\end{pmatrix}\begin{pmatrix}
        \bm b^n_j  \\
        1  \\
    \end{pmatrix}
    \right\|^2_2
    \label{eq:SensorRot}
\end{align}
has to be solved. 

In our case, this is done by reformulating the residuum in~\eqref{eq:SensorRot}. Since the second term of it is linear in $\begin{pmatrix}
    \bm R^n & \bm O^n
\end{pmatrix}$, a matrix $\mathfrak{B}_{b_j^n}\in \IR^{3\times 12}$ exists such that
\begin{align*}
    \begin{pmatrix} \bm R^n & \bm O^n
    \end{pmatrix}
    \begin{pmatrix}
        \bm b^n_j  \\
        1  \\
    \end{pmatrix} 
    =
    \mathfrak{B}_{b^n_j} 
    \begin{pmatrix}
        \text{vec}\left(\bm R^n\right) \\
        \bm O^n
    \end{pmatrix}
\end{align*}
By concatenating all $\mathfrak{B}_{b_j^n}$ vertically one can express~\eqref{eq:SensorRot} as
\begin{align*}
    \argmin_{\bm R^n , \bm O^n} 
    \left\|
    \begin{pmatrix}
        \Tilde{\bm B}_1 \\
        \vdots \\
        \Tilde{\bm B}_j
    \end{pmatrix}
    -
    \begin{pmatrix}
        \mathfrak{B}_{b^n_1} \\
        \vdots \\
        \mathfrak{B}_{b^n_J}
    \end{pmatrix}
    \begin{pmatrix}
        \text{vec}\left(\bm R^n\right) \\
        \bm O^n
    \end{pmatrix}
    \right\|^2_2
\end{align*}
which is solved by a pivoted QR factorization.

\subsubsection{Global Orientation Correction}
The coordinate system of $\Tilde{\bm B_j}$ now determines the new coordinate system of the sensor array. However, we only know the positions of the sensors $\{\bm{r}_1, \dots,\bm{r}_N \}$ in the coordinate system of the CAD model of the 3D printed sensor mount. In order to express the field using the series expansion, the next step is to specify the sensor positions in the calibration field coordinate system, resulting in corrected $t$-design positions $\{ \bm{\tilde{r}}_1, \dots, \bm{\tilde{r}}_N \}$.
Assuming that the calibration fields coordinate system is Cartesian, the correction of the $t$-design positions can be represented by a global rotation matrix $\bm M \in \textup{SO(3)}$ that rotates all positions such that
\begin{align*}
    \bm{\Tilde{r}}_n = \bm M \bm r_n,\text{ } n\in\{1, ... ,N\} .
\end{align*}
Even if the sphere coordinate system is carefully aligned with the calibration fields coordinate system, a slight sag in the mounting can cause the sensors to rotate out of their expected positions. 

For the estimation of $\bm M$, we use a measurement-based approach, combining the calibration measurement data with prior knowledge about the mounting of the sensors on the sphere's surface. 
In the case of our magnetometer array, the $z$-axes of the sensors are assumed to be perpendicular to the sphere's surface and oriented outward. This allows for an initial estimation of the sensor positions based on the orientation of their coordinate systems, which were previously derived from measurements with the calibration fields. The $z$-axis of a sensor in the coordinate system of the applied calibration fields points in the same direction as $\bm R^n_{:,z}$, which denotes the $z$ column of $\bm R^n$. We now assign to each sensor an estimated position 
\begin{align*}
    \bm{\hat{r}}_n = R \frac{\bm R^n_{:,z}}{\left\| \bm R^n_{:,z} \right\|_2} .
\end{align*}
Due to measurement inaccuracies or sensor misalignment the set of positions $\{\bm{\hat{r}}_1, \dots,\bm{\hat{r}}_N \}$ are in general no spherical $t$-design positions. 

We now search for the rotation matrix $\bm M$ that optimally aligns the $t$-design positions from the CAD model, $\{\bm{r}_1, \dots,\bm{r}_N \}$ with the estimated sensor positions,  $\{\bm{\hat{r}}_1, \dots,\bm{\hat{r}}_N \}$. Specifically, we aim to minimize the sum of squared spherical distances between these two sets of positions on the sphere $\mathbb{S}_R$ with radius $R$
\begin{align*}
    \argmin_{\bm M \in \textup{SO(3)}} \sum_{n=1}^{N}
    d_{\mathbb{S}_R}\left(
   \bm M \bm{r}_n, \bm{\hat{r}}_n
    \right)^2 ,
\end{align*}
where $d_{\mathbb{S}_R}: \mathbb{S}_R \times \mathbb{S}_R \rightarrow \IR$ is the spherical distance.

To solve the optimization problem, we parameterize $\bm M$ using a rotation angle $\alpha \in \IR$ and a rotation axis $\bm \eta \in \IR^3$ and employ the Manifolds.jl~\cite{2106.08777} package to accurately compute the spherical distances. The optimization is then performed by gradient descent using Manopt.jl~\cite{Bergmann2022}, which provides a framework for optimization on manifolds as well as a Library of optimization algorithms in Julia~\cite{Julia-2017}. As it can be seen in \cref{fig:calib}~c), the calculated rotation matrix $\bm M$ can then be applied to the sensor positions.

\subsection{Calibration Setup}
For calibration, the sphere is placed in a homogeneous  magnetic field in $J = 6$ different orientations. It is generated by a solenoid\footnote{Calibrating coil manufactured by Sekels, location: Ober-Mörlen, Germany} with an inner diameter of $\SI{30}{\cm}$. All calibration measurements were performed at a field strength of $\SI{30}{\milli\tesla}$ and lasted three minutes, providing $1800$ individual measurements that were averaged to determine all parameters of the calibration model. The relative homogeneity of the solenoid field, defined as the change in the field norm, in the spherical region is approximately $\SI{1.1e-3}{}$. Inside the volume of the Hall magnetometer, a field component perpendicular to the coil axis is generated that is approximately $\SI{1.15}{\%}$ of the nominal flux density. We consider the values to be sufficiently small to assume the field as homogeneous. Since the solenoid can only apply the field in one direction, the sphere is placed in the field in different orientations using a cubic cage mount. The right-angled structure of the mount makes it possible to record the calibration fields along the three orthogonal axes of the sphere.

\subsection{Experimental Setup}
The field measurement method using a $t$-design quadrature to discretize the field on the surface of the sphere to infer the field is already well established~\cite{Boberg2025}. We therefore compare the data from the magnetometer array with a second measurement, also using a spherical $t$-design with $t=12$ and $R=\SI{9}{\cm}$, and a spherical harmonic expansion, but recorded sequentially. The data were obtained using a single 3D Hall magnetometer (Model 460, Lake Shore, Westerville, USA) mounted on a robot that moves the sensor to $t$-design positions. With both measurement methods, a magnetic gradient field with a characteristic field-free point (FFP) and a gradient strength of $\SI{0.22}{\tesla\per\meter}$ in $y$-direction and $\SI{0.11}{\tesla\per\meter}$ in $x$ and $z$ direction is analyzed~\cite{thieben_system_2024}. For a meaningful comparison, it is important to ensure that we measure the same volume and in the same orientation. By aligning the FFPs to the center of each measurement volume it is guaranteed that both approaches capture the same region. To achieve proper orientation, the magnetometer array is initially rotated as precisely as possible to align its axes with those of the single sensor measurement. Any remaining misorientation is corrected by applying an additional optimization in a post-processing step by using a simplex algorithm, which rotates the array’s field to best match that of the single sensor measurement. Additionally, the sphere’s radius is included in the optimization to compensate for minor discrepancies in size due to possible inaccuracies in the 3D printing process. To obtain statistical values for the deviation of the two field measurement methods, the field within the sphere is sampled Cartesian with a grid spacing of \SI{0.5}{\mm}.

In a second experiment, a soft iron coil array~\cite{foerger_flexible_2023} is used to generate a time-dependent gradient field that provides a dynamic FFP. The field is activated for about $\SI{5}{\s}$ and during this time the FFP performs a continuous oval periodic motion with varying speed within the spherical shape of the magnetometer array. The trajectory of the FFP is recorded live at $\SI{10}{\Hz}$ using the magnetometer array. The measured data are serially transferred to a computer via the USB interface, where the field inside the sphere is calculated from the field values at the $t$-design positions.

\section{Results}
\begin{figure}[]
    \centering
    \input{tikz/5-results_FFPTrajectory}
    \caption{Measured positions of the field-free point\,(FFP) along a closed, oval trajectory with a duration of \SI{5}{\second}. The field is acquired with \SI{10}{\hertz}. Each circle represents the FFP position at a consecutive point in time, with the color scheme illustrating the chronological sequence for enhanced visualization. The projections onto the coordinate planes are shown in gray.}
    \label{fig:FFPTrajectory}
\end{figure}
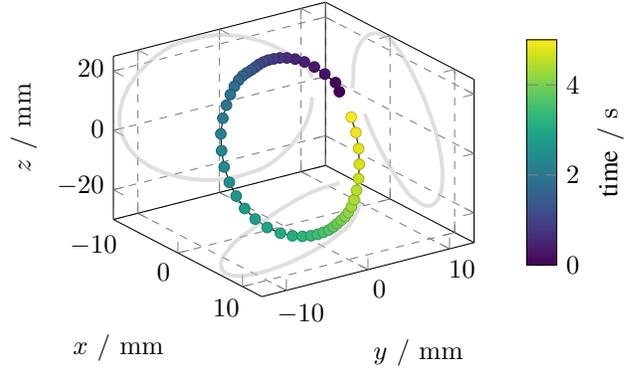
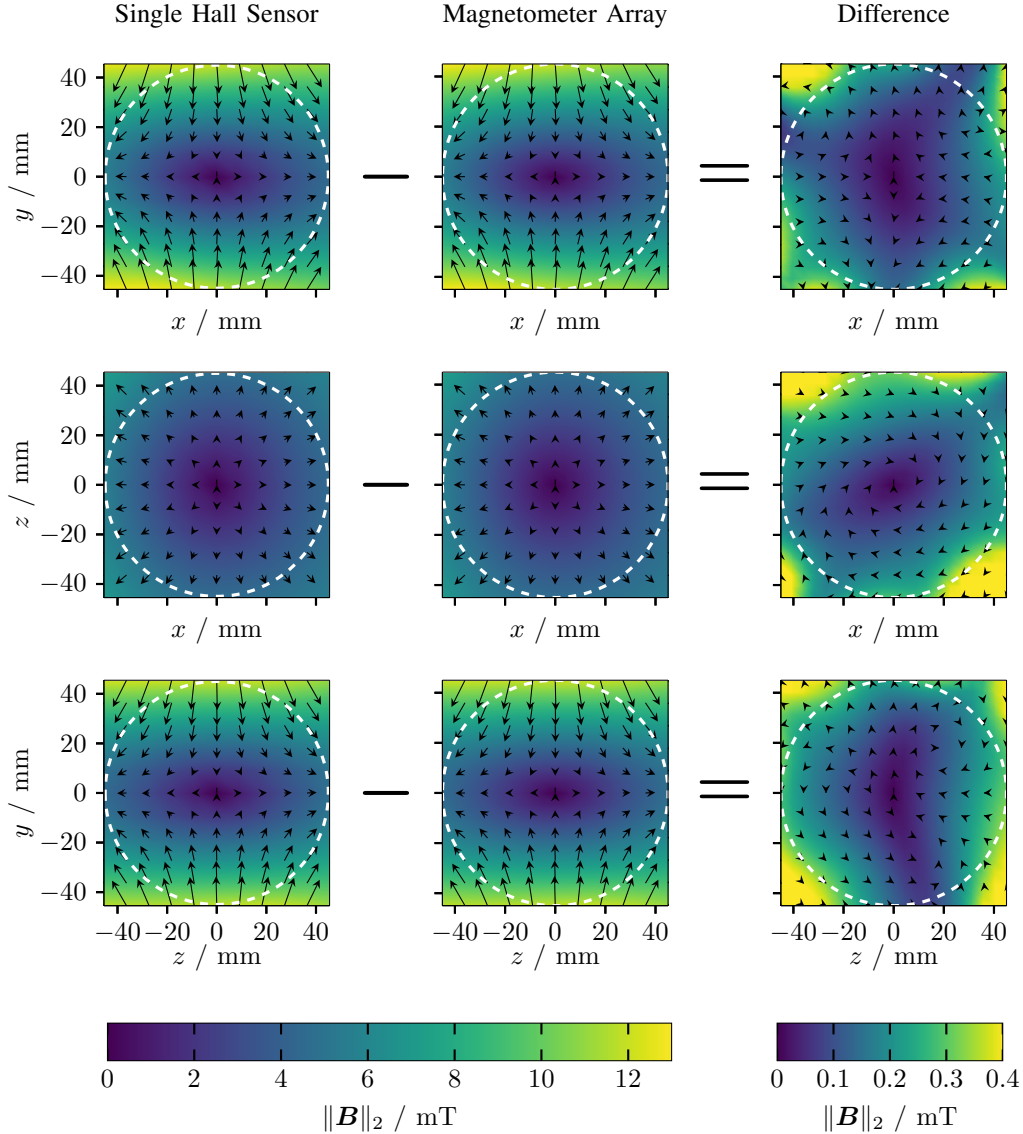
\begin{figure*}[h]
    \centering
    \input{tikz/4-results_fieldPlot.tex}
    \caption{Comparison of field measurements from the single sensor approach and the magnetometer array across three orthogonal planes. The examined field is a gradient field with a field-free point at the center and a gradient strength of \SI{0.22}{\tesla\per\meter} in $y$-direction and $\SI{0.11}{\tesla\per\meter}$ in $x$ and $z$ direction. The left column displays the single sensor data, the middle column the magnetometer array data, and the right column the magnitude and direction of the difference vector between the two measurements. Only inside the sphere's boundary (white circle), the fields series expansion is valid. Values outside are to be ignored.}
    \label{fig:PlanesComaprison}
\end{figure*}

We first consider the results for the measurement of the static gradient field. Both, the single-sensor and the magnetometer array yield very similar magnetic field measurements, but with very different scan times. While sequential measurement with a single sensor and a robot takes about $\SI{3}{\min}$, the magnetometer array provides equivalent data in only $\SI{100}{\milli\second}$.

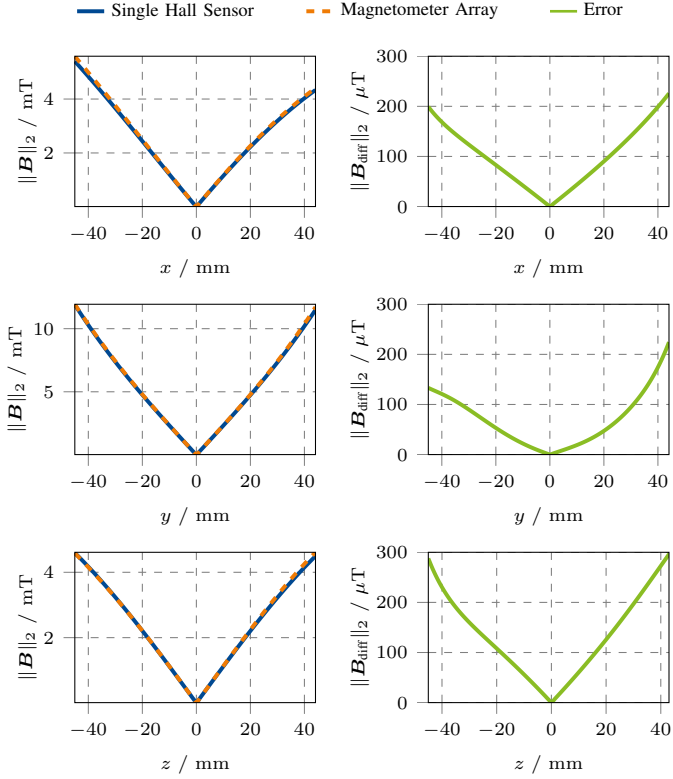
\begin{figure}
    \centering
    \input{tikz/3-results_fieldAccuracy_upright}
    \caption{Comparison of field measurements using the single-sensor approach and the magnetometer array along the three coordinate axes. The left diagrams depict the absolute field values for both methods, while the right diagrams illustrate the magnitude of the difference field between the two measurements.}
    \label{fig:AxesComaprison}
\end{figure}

The optimization in the post-processing step to align the two field measurements with each other resulted in a tilt angle of $\SI{1.8}{\degree}$. The optimized radius of the sphere is $\SI{45.41}{\mm}$.  After correcting for the misalignment and scaling of the radius, the average norm of the difference field of the two fields within the sphere is \SI{154\pm66}{\micro\tesla} (mean $\pm$ standard deviation), while the maximum is \SI{403}{\micro\tesla}. In relation to the maximum magnitude of the field within the sphere of \SI{12.0}{\milli\tesla}, this results in an average deviation of about $1\%$ and a maximum deviation of about $4\%$.  A plot of the field obtained by the two measurement methods as a plot of the difference field on the cross-sectional planes is shown in \cref{fig:PlanesComaprison}. A direct comparison of the two measurement methods is presented in \cref{fig:AxesComaprison}. The figure shows the measured absolute field values (left column) and the absolute value of the difference field (right column) along the coordinate axes from \cref{fig:PlanesComaprison}. For both images it can be seen that the largest absolute deviations are observed at the edge of the sphere. Note that the approximation of the field by a spherical harmonic expansion does not provide information about the field outside this sphere~\cite{Boberg2025}, as indicated by the white dashed line.

\cref{fig:FFPTrajectory} shows the recorded trajectory of an FFP moving inside the spherical volume. Within \SI{5}{\s}, $50$ magnetic field shots were taken. The trajectory is oval in shape and the speed of the FFP varies according to its position, as can be seen from the different distances between the recorded points along the trajectory. 

\section{Discussion}
The results demonstrate that the calculated polynomial is an accurate representation of the magnetic field. Measurement accuracy depends on both the complexity of the spatial profile of the magnetic fields being analyzed and the calibration process. In particular, fields that can be well approximated by a polynomial of degree $6$ can be accurately measured. Since the analyzed field fulfills this requirement, we can focus here on errors originating from other sources of uncertainty.

A crucial requirement for the calibration method is that the calibration fields $\Tilde{\bm B}_j$ must be highly homogeneous, and their amplitudes must be determined with high accuracy. In our study, the expected errors due to minor inhomogeneities in the calibration fields are an order of magnitude smaller than the observed error. Therefore, additional sources of error must be contributing.
Another limiting factor is the susceptibility of Hall magnetometers to temporal drifts, which can be induced by temperature fluctuations~\cite{9296742}. So far, temperature measurements within the array have not been used to model potential drift effects. Additionally, there is some uncertainty regarding the shape of the 3D printed sensor holder, which is not perfectly spherical. For a field gradient of \SI{0.2}{\tesla\per\meter}, even a \SI{1}{mm} radial displacement of a sensor would result in a field variation of \SI{200}{\micro\tesla}. If multiple sensors systematically deviate from the expected radial distance in a non-spherical manner, this can lead to distortions in the measured field gradients that cannot be captured by optimizing the sphere radius, potentially affecting the accuracy of the field reconstructions.

Another source of uncertainty arises from the fact that the earth's magnetic field has not yet been incorporated into the calibration, leading to an additional measurement deviation of approximately \SI{48}{\micro\tesla}. Considering all these factors, the observed average deviation of \SI{131}{\micro\tesla} between the two measurement methods is in good agreement with the discussed sources of errors, confirming the consistency of the measurements.

Comparing the measurement time of the sequential and parallel method for a spherical volume with a diameter of \SI{9}{\cm}, the parallel measurement leads to a speedup by a factor of about $1800$. With this high frame rate, the movement of an FFP in a field generator could be recorded live. The oval trajectory and varying speed are expected, as the circular motion was only roughly modeled via the coil currents without a precisely controlled field sequence. The strong field inhomogeneity and varying gradient strengths throughout the volume further prevent a uniform FFP movement.

\section{Conclusion}
The method of measuring magnetic fields at spherical $t$-design positions and reconstructing the field within the sphere's volume using a spherical harmonic expansion was successfully applied to a magnetometer array. The presented calibration methods enable monitoring of a magnetic field that can be well represented by a polynomial of degree 6, with a temporal resolution of \SI{10}{\Hz} and an accuracy in the \SI{100}{\micro\tesla} range. This capability opens up entirely new application scenarios where the spatio-temporal distribution of magnetic fields is of interest.

The magnetometer array is especially useful in measurement scenarios where a large number of volumetric magnetic field measurements need to be carried out. For example, the shimming process in MRI systems involves repeated magnetic field measurements to assess and correct for field inhomogeneities~\cite{wachowicz_evaluation_2014}. This process, which typically involves iterative adjustments, would particularly benefit from a fast and volumetric field measurement approach like the proposed magnetometer array, enabling real-time feedback and improved optimization.
Another example is the operation of magnetic field generators for Magnetic Particle Imaging Systems or Magnetic Manipulation Systems, where numerous magnetic field calibration measurements are required~\cite{DesignAndOpt,yang_magnetic_2020}. Additionally, the measurement method inherently offers good scalability, allowing for its application to larger volumes and the integration of additional sensors.

The ability to track an FFP's position in real time highlights the potential of the magnetometer array for dynamic field measurements, making it particularly relevant for applications in MPI and other imaging techniques. A precise knowledge of the field and its change over time allows a higher accuracy for the calculation of the imaging area, resolution~\cite{knopp_magnetic_2017} and model-based reconstructions~\cite{ModelBasedKnopp,thieben_experimental_2024}. Especially for the planning and characterization of general multi-patch sequences \cite{DesignAndOpt}, a high frame rate of the field measurement device is crucial.  The device can be used to directly map how the selection field changes during a sequence under the influence of the field generating coils and additional eddy currents. The measurement effort with the sequential single sensor approach would be significantly greater here, as many individual time-triggered measurements would have to be carried out during the sequence. 

\section{Outlook}
Future work will focus on refining the calibration process by incorporating the earth's magnetic field and improving sensor positioning on the Hall magnetometer holder. For direct verification, additional sensors can be placed within the sphere. Our current sampling rate of \SI{10}{\Hz} does not yet reach hardware limitations, suggesting that an increase by a factor of ten is feasible. Furthermore, the integration of small receiver coils, which are fed via multiplexers to a suitable analogue-to-digital converter and digitized, is a promising approach for AC field measurements. Another interesting direction is extending this technique to ellipsoidal regions, using ellipsoidal harmonics as the field expansion basis~\cite{scheffler_ellipsoidal_2024,Scheffler2025Efficient}. Since magnetic fields are often generated and analyzed in non-spherical regions, an ellipsoidal approach could better adapt to the actual measurement volume. For instance, tomographic imaging systems typically operate in cylindrical regions, where an ellipsoidal measurement device could provide improved spatial coverage and more accurate field characterization.

\section*{Declarations}
\subsection*{Conflict of Interest}
The authors declare no competing interests.


\bibliographystyle{IEEEtran}

\bibliography{ref}

\end{document}

%% file: math-definitions.tex
\usepackage{amsmath}
\usepackage{amsfonts}
\usepackage{amssymb}
\usepackage{nicefrac}
\usepackage{bbm} 
\usepackage{bm} 

\usepackage[all]{xy}

\newcommand{\IR}{\mathbb{R}}		

\renewcommand{\epsilon}{\varepsilon}				
\renewcommand{\theta}{\vartheta}                  	
\renewcommand{\phi}{\varphi}						

\newcommand{\set}[2][]{\left\{ #2 
                        \ifthenelse{\equal{#1}{}}{}{: #1}
                        \right\}}				        

\DeclareMathOperator*{\argmin}{arg\,min}

\usepackage[exponent-product = \cdot,separate-uncertainty=true]{siunitx}
\DeclareSIUnit{\mup}{\text{$\mu_0$}}

\usepackage{amsthm}
\theoremstyle{plain} 

\theoremstyle{definition} 

\theoremstyle{remark} 

\usepackage{tikz}
\usepackage{tikz-3dplot}
\usepackage{pgfplots}
\usepackage{pgfplotstable}

\pgfplotsset{compat=1.8}

\usetikzlibrary{shapes,shapes.multipart,positioning,shapes.arrows, arrows.meta, decorations.pathmorphing, calc,fadings}
\usepgfplotslibrary{units,statistics,groupplots}

\pgfplotsset{
    unit markings=slash space,
    /pgfplots/xbar/.style={
    /pgf/bar shift={-0.5*(\numplotsofactualtype*\pgfplotbarwidth + (\numplotsofactualtype-1)*#1) + (.5+\plotnumofactualtype)*\pgfplotbarwidth + \plotnumofactualtype*#1},
    },
}
\pgfplotsset{select coords between index/.style 2 args={
    x filter/.code={
        \ifnum\coordindex<#1\def\pgfmathresult{}\fi
        \ifnum\coordindex>#2\def\pgfmathresult{}\fi
    }
}}

\usepackage{color}
\definecolor{ibilight}{RGB}{193,216,237}
\definecolor{ibidark}{RGB}{0,73,146}	
\definecolor{uke2}{RGB}{170,156,143} 	
\definecolor{uke3}{RGB}{87,87,86}		
\definecolor{ukesec1}{RGB}{255,223,0}	
\definecolor{ukesec2}{RGB}{239,123,5}	
\definecolor{ukesec3}{RGB}{104,195,205}	
\definecolor{ukesec4}{RGB}{138,189,36}	
\definecolor{ukesec5}{RGB}{178,34,41}	
\definecolor{tuhh}{RGB}{45,198,214}     
\definecolor{ibidarkBG}{RGB}{227,229,242}   
\definecolor{uke2BG}{RGB}{233,228,225} 	    
\definecolor{uke3BG}{RGB}{230,231,232}	    
\definecolor{ukesec1BG}{RGB}{255,243,190}   
\definecolor{ukesec2BG}{RGB}{254,232,212}   
\definecolor{ukesec3BG}{RGB}{222,241,241}   
\definecolor{ukesec4BG}{RGB}{233,243,222}   
\definecolor{ukesec5BG}{RGB}{244,230,225}   

\usepackage[capitalise]{cleveref} 

\makeatletter
\newif\iftikz@shading@path

\tikzset{
    shading xsep/.store in=\tikz@pathshadingxsep,
    shading ysep/.store in=\tikz@pathshadingysep,
    shading sep/.style={shading xsep=#1, shading ysep=#1},
    shading sep=0.0cm,
}

\def\tikz@shadepath#1{%
    \iftikz@shading@path%
    \else%
        \tikz@shading@pathtrue%
        \pgfgetpath\tikz@currentshadingpath%
        \begingroup%
            \pgfsys@beginscope
            \tikzset{#1}%
            \xdef\tikz@tmp{\noexpand\def\noexpand\tikz@pathshadingxsep{\tikz@pathshadingxsep}%
                \noexpand\def\noexpand\tikz@pathshadingysep{\tikz@pathshadingysep}}%
            \pgfsys@endscope%
        \endgroup
        \tikz@tmp%
        \pgfextract@process\pgf@shadingpath@southwest{\pgfpointadd{\pgfqpoint{\pgf@pathminx}{\pgf@pathminy}}%
            {\pgfpoint{-\tikz@pathshadingxsep}{-\tikz@pathshadingysep}}}
        \pgfextract@process\pgf@shadingpath@northeast{\pgfpointadd{\pgfqpoint{\pgf@pathmaxx}{\pgf@pathmaxy}}%
            {\pgfpoint{\tikz@pathshadingxsep}{\tikz@pathshadingysep}}}%
        \pgfsetpath\pgfutil@empty%
        \let\tikz@options@saved=\tikz@options%
        \let\tikz@mode@saved=\tikz@mode%
        \let\tikz@options=\pgfutil@empty%
        \let\tikz@mode=\pgfutil@empty%
        \tikz@addoption{%
            \pgfinterruptpath%
            \pgfinterruptpicture%
                \begin{tikzfadingfrompicture}[name=.]
                \pgfscope%
                    \tikzset{shade path/.style=}
                    \path \pgfextra{%
                        \pgfsetpath\tikz@currentshadingpath%
                        \pgf@shadingpath@southwest
                        \expandafter\pgf@protocolsizes{\the\pgf@x}{\the\pgf@y}%
                        \pgf@shadingpath@northeast%
                        \expandafter\pgf@protocolsizes{\the\pgf@x}{\the\pgf@y}%
                        \let\tikz@options=\tikz@options@saved%
                        \let\tikz@mode=\tikz@mode@saved%
                    };
                    \xdef\pgf@shadingboundingbox@southwest{\noexpand\pgfqpoint{\the\pgf@picminx}{\the\pgf@picminy}}%
                    \xdef\pgf@shadingboundingbox@northeast{\noexpand\pgfqpoint{\the\pgf@picmaxx}{\the\pgf@picmaxy}}%
                    \endpgfscope
                \end{tikzfadingfrompicture}%
            \endpgfinterruptpicture%
            \endpgfinterruptpath%
            \pgftransformreset%
            \pgfpathrectanglecorners{\pgf@shadingboundingbox@southwest}{\pgf@shadingboundingbox@northeast}%
            \let\tikz@path@picture=\pgfutil@empty%
            \tikz@mode@fillfalse%
            \tikz@mode@drawfalse%
            \tikz@mode@doublefalse%
            \tikz@mode@clipfalse%
            \tikz@mode@boundaryfalse%
            \tikz@mode@fade@pathfalse%
            \tikz@mode@fade@scopefalse%
            \tikzset{#1}%
            \tikz@mode%
            \def\tikz@path@fading{.}%
            \tikz@mode@fade@pathtrue%
            \tikz@fade@adjustfalse%
            \pgfpointscale{0.5}{\pgfpointadd{\pgf@shadingboundingbox@southwest}{\pgf@shadingboundingbox@northeast}}%
            \edef\tikz@fade@transform{shift={(\the\pgf@x,\the\pgf@y)}}%
            \pgfsetfading{\tikz@path@fading}{\tikz@do@fade@transform}%
            \tikz@mode@fade@pathfalse%
        }%
    \fi%
}
\tikzset{
    shade path/.code={%
        \tikz@addmode{\tikz@shadepath{#1}}%
    }
}
\makeatother 

%% file: acronyms.tex
\usepackage[nolist,nohyperlinks]{acronym}

\begin{acronym}[ECU]

    \acro{MRI}[MRI]{magnetic resonance imaging}
    \acro{MPI}[MPI]{magnetic particle imaging}
    
    \acro{SPIOs}[SPIOs]{superparamagnetic iron oxide nanoparticles}

    \acro{FFP}[FFP]{field-free-point}
    \acro{FFL}[FFL]{field-free-line}
    \acro{FOV}[FOV]{field-of-view}
    \acro{LFR}[LFR]{low-field-region}
    
    \acro{SF}[SF]{selection field}
    \acro{DF}[DF]{drive field}
    \acro{FF}[FF]{focus field}
    
    \acro{ADC}[ADC]{analog-to-digital converter}

\end{acronym}

%% file: tikz/1-intro_motivation.tex
\graphicspath{{images/}}

\def\size{1cm}
\def\sizeFieldPlot{3.1cm}
\def\nodeXDistance{0.8cm}
\def\nodeYDistance{1cm}
\def\shift{1cm}
\def\sizeNote{1cm}
\def\sizeNoteCalib{1cm}
\centering
\begin{center}
\begin{tikzpicture}[node distance = 1cm and 1cm]
\tikzstyle{ar} = [-{Triangle[width=10pt,length=10pt]}, line width=5pt, ibidark]

    \tikzset{
        >={stealth},
        notes/.style = {
        text width = \sizeNote,
        font=\small,
        text depth = 2.5 cm,
        },
        node distance=\nodeYDistance and \nodeXDistance,
    }


\node[ inner sep=0pt,xshift=2*\nodeXDistance] (SelFieldGrid) {\includegraphics[trim={1.135cm 1.005cm 1cm 0.8cm 0},clip,height=\sizeFieldPlot]{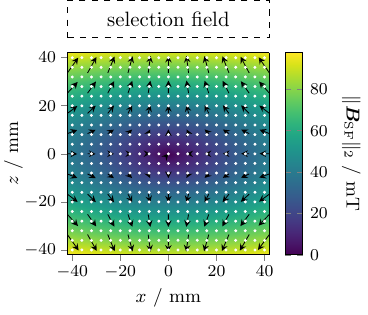}};
\node[anchor=center,inner sep=0pt, xshift=-\nodeXDistance,rotate=0,right = of SelFieldGrid] (SelFieldGridCB) {\rotatebox{90}{\small $\left\lvert \bm B \right\rvert$}};

\node[right = of SelFieldGrid, inner sep=0pt] (SelFieldtDesign) {\includegraphics[trim={1.135cm 1.005cm 1cm 0.8cm 0},clip,height=\sizeFieldPlot]{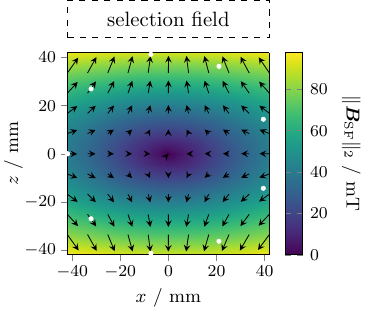}};
\node[anchor=center,inner sep=0pt,xshift=-\nodeXDistance,rotate=0,right = of SelFieldtDesign] (SelFieldtDesignCB) {\rotatebox{90}{\small $\left\lvert \bm B \right\rvert$}};

\node[right = of SelFieldtDesign, inner sep=0pt] (HallSensorSphere) {\includegraphics[trim={0cm 0cm 0cm 0cm},clip,height=\sizeFieldPlot]{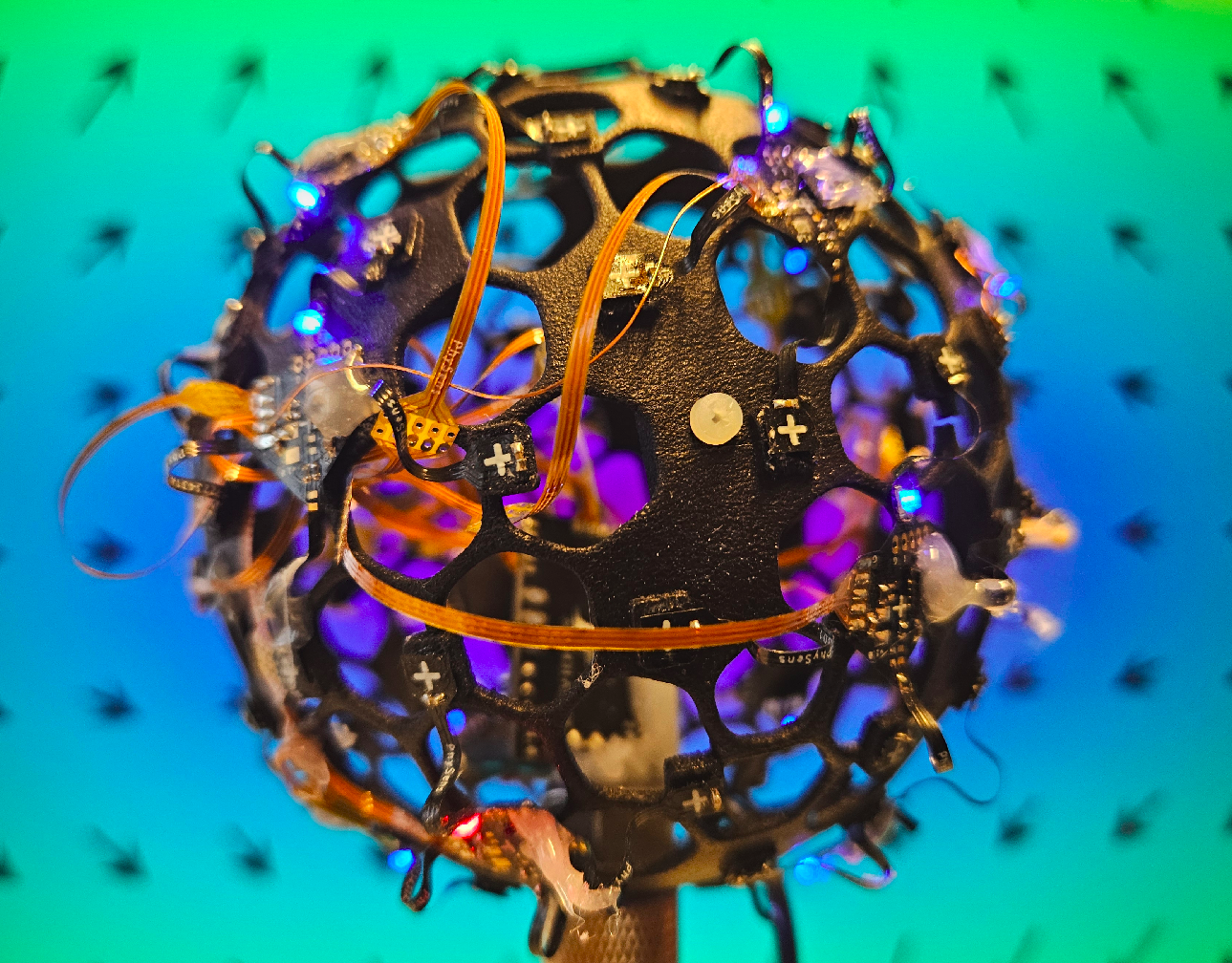}};

\node[above = 0 mm of SelFieldGrid, text width = 30mm, align=center,,xshift = -2.5mm] (){\footnotesize Grid Sampling};
\node[above = 0 mm of SelFieldtDesign, text width = 30mm, align=center,xshift = -2.5mm] (){\footnotesize t-Design Sampling};
\node[above = 0 mm of HallSensorSphere, text width = 30mm, align=center] (){\footnotesize Magnetometer Array};

\path[draw=transparent!0,line width=0.7cm, line cap=butt, -{Stealth[length=9mm, width=13mm]}, 
    shade path={ left color=white, right color=ibidark}] 
        ($(SelFieldGrid.south west)-(0,1)$) -- ($(HallSensorSphere.south east)-(0.1,1)$);
          
\node[yshift=-1cm, inner sep=0pt,text width=\sizeFieldPlot+1cm,align=center] (TimingMagSphere) at (HallSensorSphere.south) {\centering {\textcolor{white}{$T_{\textup{meas}} = \SI{0.1}{\s}$}}};
\node[below = 0.15cm of TimingMagSphere]{\small \textit{parallel}};

\node[yshift=-1cm, inner sep=0pt,text width=\sizeFieldPlot+1cm,align=center] (TimingSelFieldGrid) at (SelFieldGrid.south) {\centering {\textcolor{black}{$T_{\textup{meas}} \gg \SI{3}{\min}$}}};
\node[below = 0.15cm of TimingSelFieldGrid](seq1Node){\small \textit{sequential}};

\node[yshift=-1cm, inner sep=0pt,text width=\sizeFieldPlot+1cm,align=center] (TimingSelFieldtDesign) at (SelFieldtDesign.south) {\centering {\textcolor{black}{$T_{\textup{meas}} \approx \SI{3}{\min}$}}};
\node[below = 0.15cm of TimingSelFieldtDesign](seq2Node){\small \textit{sequential}};

\node (b) at ($(seq1Node)!0.5!(seq2Node)$) {};
    \node[inner sep=0pt, yshift=0mm] at (b)(Isel) {\includegraphics[height=1.2cm]{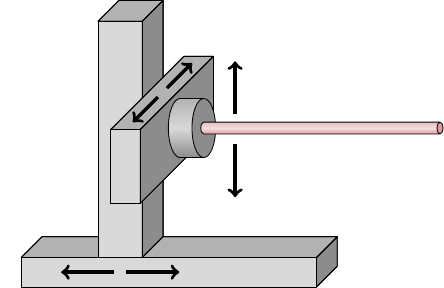}};

\node[left = -18mm of TimingSelFieldGrid.west, inner sep=0pt,text width=\sizeFieldPlot+1cm,align=center] (turtle) {\includegraphics[scale = 0.027]{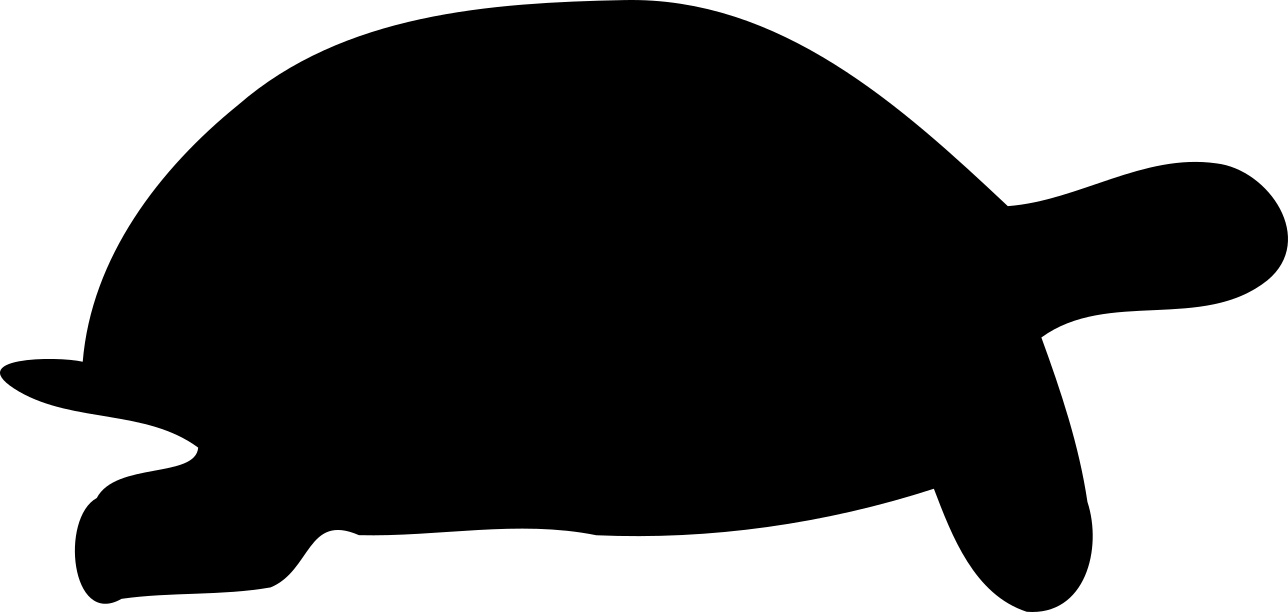}};
\node[right = -16mm of TimingMagSphere.east, inner sep=0pt,text width=\sizeFieldPlot+1cm,align=center] (turtle) {\includegraphics[scale = 0.022]{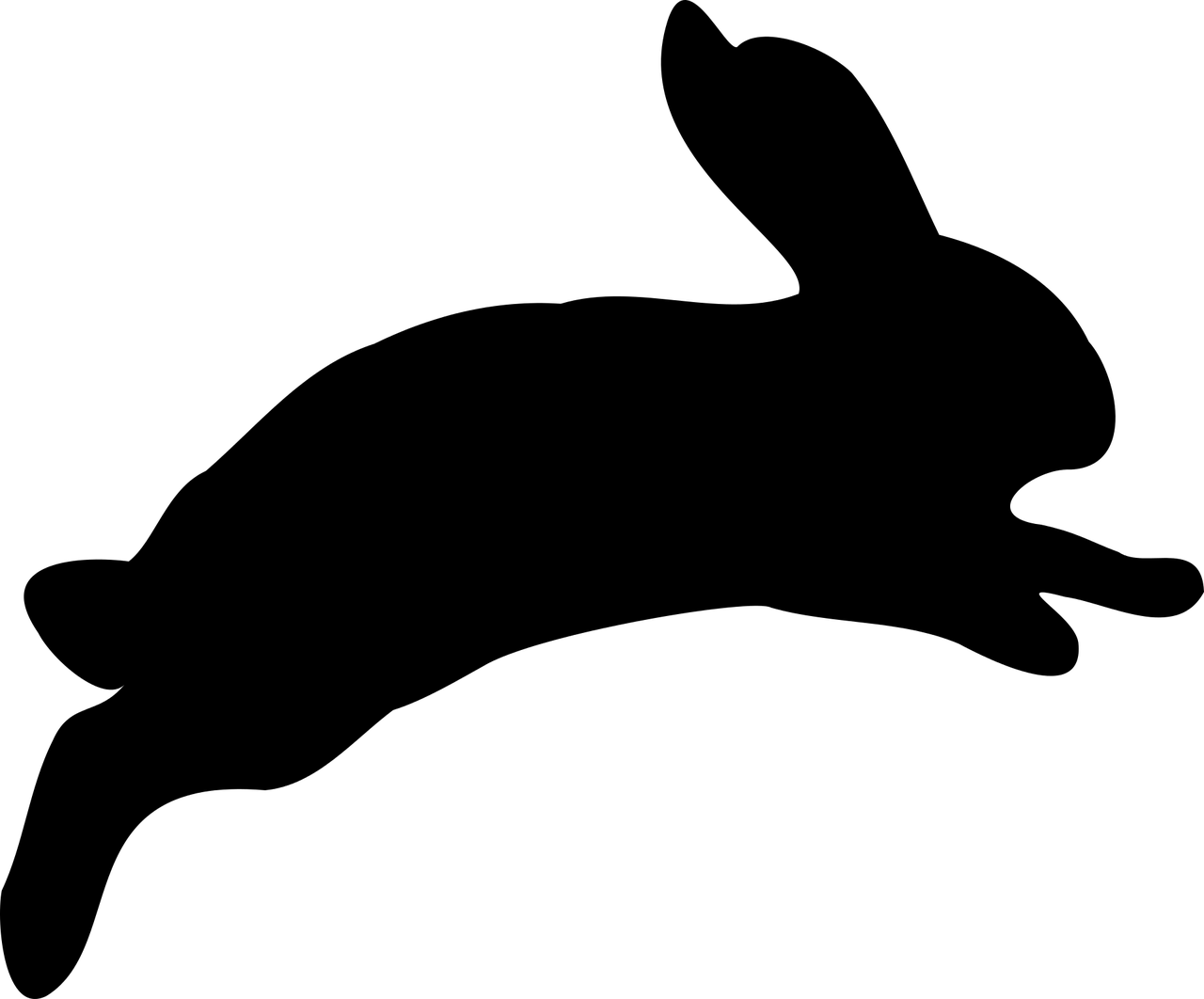}};


\node[xshift=-0.3cm] at (SelFieldGrid.west) (endArrow){};


\end{tikzpicture}
\end{center}

%% file: tikz/2-theory_calibration.tex

\tikzset{>={stealth}}
\tikzstyle{ar} = [-{Triangle[width=10pt,length=6pt]}, line width=4pt, ukesec3]

    \tikzset{
        >={stealth},
        notes/.style = {
        text width = \sizeNote,
        font=\small,
        text depth = 2.5 cm,
        },
        capt/.style = {
                    text width = 20cm,
                    text centeukesec2,
                    },
        node distance=\nodeYDistance and \nodeXDistance,
    }

\def\r{1.5}
\def\numberOfAngles{7}
\def\lengthCoordinateArrow{0.4}
\def\rotationOfMagSphere{20}
\def\opacity{0.35}
\def\coordinateSystemDisturbanceZDirX{{0.0667, 0.02025, 0.0566, 0.0634, 0.0154, 0.0784, 0.0803}}
\def\coordinateSystemDisturbanceZDirY{{0.0041, 0.087782, 0.07148839458111712, 0.05700, 0.02796, 0.08564608810067957, 0.09509}}
\def\coordinateSystemDisturbanceYDirX{{0.02933, 0.02264, 0.09470521084974409, 0.034933, 0.051720, 0.069985, 0.05105}}
\def\coordinateSystemDisturbanceYDirY{{0.091203, 0.03370, 0.0416419, 0.050164, 0.097978, 0.06866, 0.0045328}}

\begin{tikzpicture}

  \shade[ball color = gray!40, opacity = \opacity] (0,0) circle (\r);
  \draw (0,0) circle (\r);
  \draw (-\r,0) arc (180:360:1.5 and 0.3); 
  \draw[dashed] (\r,0) arc (0:180:1.5 and 0.3); 
\draw[<->] (0,0) -- (0,\r) node[midway, right] {\footnotesize \( R \)};
\foreach \i in {0,...,\numexpr\numberOfAngles-1\relax}{
  \draw[->,ukesec2,line width=1pt] ({\r*sin(\i*360/\numberOfAngles+\rotationOfMagSphere)},{\r*cos(\i*360/\numberOfAngles+\rotationOfMagSphere)})--({(\lengthCoordinateArrow/\r+1)*\r*sin(\i*360/\numberOfAngles+\rotationOfMagSphere)+\coordinateSystemDisturbanceZDirX[\i]},{(\lengthCoordinateArrow/\r+1)*\r*cos(\i*360/\numberOfAngles+\rotationOfMagSphere)+\coordinateSystemDisturbanceZDirY[\i]});
  }
  
\foreach \i in {0,...,\numexpr\numberOfAngles-1\relax}{
  \draw[->,ibidark,line width=1pt] ({\r*sin(\i*360/\numberOfAngles+\rotationOfMagSphere)},{\r*cos(\i*360/\numberOfAngles+\rotationOfMagSphere)})--({\r*sin(\i*360/\numberOfAngles+\rotationOfMagSphere)+\lengthCoordinateArrow*cos(\i*360/\numberOfAngles+\rotationOfMagSphere)+\coordinateSystemDisturbanceYDirX[\i]},{\r*cos(\i*360/\numberOfAngles+\rotationOfMagSphere)-\lengthCoordinateArrow*sin(\i*360/\numberOfAngles+\rotationOfMagSphere)+\coordinateSystemDisturbanceYDirY[\i]});
  }
  

\def\shiftX{5}

  \shade[ball color = gray!40, opacity = \opacity] (\shiftX,0) circle (\r);
  \draw (\shiftX,0) circle (\r);
  \draw (-\r+\shiftX,0) arc (180:360:1.5 and 0.3); 
  \draw[dashed] (\r+\shiftX,0) arc (0:180:1.5 and 0.3); 

\foreach \i in {0,...,\numexpr\numberOfAngles-1\relax}{
  \draw[->,ukesec2,line width=1pt, opacity = \opacity] ({\r*sin(\i*360/\numberOfAngles+\rotationOfMagSphere)+\shiftX},{\r*cos(\i*360/\numberOfAngles+\rotationOfMagSphere)})--({(\lengthCoordinateArrow/\r+1)*\r*sin(\i*360/\numberOfAngles+\rotationOfMagSphere)+\coordinateSystemDisturbanceZDirX[\i]+\shiftX},{(\lengthCoordinateArrow/\r+1)*\r*cos(\i*360/\numberOfAngles+\rotationOfMagSphere)+\coordinateSystemDisturbanceZDirY[\i]});
  }
  
\foreach \i in {0,...,\numexpr\numberOfAngles-1\relax}{
  \draw[->,ibidark,line width=1pt, opacity = \opacity] ({\r*sin(\i*360/\numberOfAngles+\rotationOfMagSphere)+\shiftX},{\r*cos(\i*360/\numberOfAngles+\rotationOfMagSphere)})--({\r*sin(\i*360/\numberOfAngles+\rotationOfMagSphere)+\lengthCoordinateArrow*cos(\i*360/\numberOfAngles+\rotationOfMagSphere)+\coordinateSystemDisturbanceYDirX[\i]+\shiftX},{\r*cos(\i*360/\numberOfAngles+\rotationOfMagSphere)-\lengthCoordinateArrow*sin(\i*360/\numberOfAngles+\rotationOfMagSphere)+\coordinateSystemDisturbanceYDirY[\i]});
  }

\foreach \i in {0,...,\numexpr\numberOfAngles-1\relax}{
  \draw[->,ukesec2,line width=1pt] ({\r*sin(\i*360/\numberOfAngles+\rotationOfMagSphere)+\shiftX},{\r*cos(\i*360/\numberOfAngles+\rotationOfMagSphere)})--({\r*sin(\i*360/\numberOfAngles+\rotationOfMagSphere)+\shiftX},{\r*cos(\i*360/\numberOfAngles+\rotationOfMagSphere)+\lengthCoordinateArrow});
  }
  
\foreach \i in {0,...,\numexpr\numberOfAngles-1\relax}{
  \draw[->,ibidark,line width=1pt] ({\r*sin(\i*360/\numberOfAngles+\rotationOfMagSphere)+\shiftX},{\r*cos(\i*360/\numberOfAngles+\rotationOfMagSphere)})--({\r*sin(\i*360/\numberOfAngles+\rotationOfMagSphere)+\shiftX+\lengthCoordinateArrow},{\r*cos(\i*360/\numberOfAngles+\rotationOfMagSphere)});
  }

\def\shiftX{10}

  \shade[ball color = gray!40, opacity = \opacity] (\shiftX,0) circle (\r);
  \draw (\shiftX,0) circle (\r);
  \draw (-\r+\shiftX,0) arc (180:360:1.5 and 0.3); 
  \draw[dashed,black!50] (\r+\shiftX,0) arc (0:180:1.5 and 0.3); 
  \draw[line width = 2pt,ukesec5!80!black] (\shiftX,-0.3) -- (\shiftX,-0.7);
  \draw[line width = 2pt,ukesec5!60!white] (\shiftX,-0.3) -- (\shiftX,0.3);
  \draw[line width = 2pt,ukesec5!20!white] (\shiftX,0.3) -- (\shiftX,0.4);
  \fill[ukesec5!80!white] (\shiftX,-0.7) circle (1pt);
  \fill[ukesec5!80!black] (\shiftX,-0.3) circle (1pt);
  \fill[ukesec5!60!white] (\shiftX,0.3) circle (1pt);
  \fill[ukesec5!20!white] (\shiftX-1,0.4) -- (\shiftX+0.03527778,0.4) arc [start angle=0, end angle=180, radius=1pt] -- cycle;
  \node at (\shiftX+0.2,0) {\footnotesize $\bm \eta$};
  
  \draw[line width = 0.5pt] (\shiftX,0) -- ({\r*sin(6*360/\numberOfAngles+\rotationOfMagSphere)+\shiftX},{\r*cos(6*360/\numberOfAngles+\rotationOfMagSphere)});
  \draw[line width = 0.5pt] (\shiftX,0) -- ({\r*sin(6*360/\numberOfAngles)+\shiftX},{\r*cos(6*360/\numberOfAngles)});

  \path[fill=black!40,draw] (\shiftX,0) -- ({\r*sin(6*360/\numberOfAngles+\rotationOfMagSphere)+\shiftX},{\r*cos(6*360/\numberOfAngles+\rotationOfMagSphere)}) arc[start angle=90+31, end angle=110+31, radius=\r] -- (\shiftX,0) -- cycle;

  \fill (\shiftX,0) circle (1pt);
  \begin{scope}[yscale=-1]
      \draw[->,thick, black,line width = 0.75pt] (\shiftX,0.6) +(40:0.15cm) arc(40:-220:0.15cm) -- ++(50:2pt); 
  \end{scope}
  \node at (\shiftX-0.65,0.75) {\footnotesize $\alpha$};
  

\foreach \i in {0,...,\numexpr\numberOfAngles-1\relax}{
  \draw[->,ukesec2,line width=1pt, opacity=0.3] ({\r*sin(\i*360/\numberOfAngles+\rotationOfMagSphere)+\shiftX},{\r*cos(\i*360/\numberOfAngles+\rotationOfMagSphere)})--({\r*sin(\i*360/\numberOfAngles+\rotationOfMagSphere)+\shiftX},{\r*cos(\i*360/\numberOfAngles+\rotationOfMagSphere)+\lengthCoordinateArrow});
  }
  
\foreach \i in {0,...,\numexpr\numberOfAngles-1\relax}{
  \draw[->,ibidark,line width=1pt, opacity=0.3] ({\r*sin(\i*360/\numberOfAngles+\rotationOfMagSphere)+\shiftX},{\r*cos(\i*360/\numberOfAngles+\rotationOfMagSphere)})--({\r*sin(\i*360/\numberOfAngles+\rotationOfMagSphere)+\shiftX+\lengthCoordinateArrow},{\r*cos(\i*360/\numberOfAngles+\rotationOfMagSphere)});
  }

\def\rotationOfMagSphere{0}
\foreach \i in {0,...,\numexpr\numberOfAngles-1\relax}{
  \draw[->,ukesec2,line width=1pt] ({\r*sin(\i*360/\numberOfAngles+\rotationOfMagSphere)+\shiftX},{\r*cos(\i*360/\numberOfAngles+\rotationOfMagSphere)})--({\r*sin(\i*360/\numberOfAngles+\rotationOfMagSphere)+\shiftX},{\r*cos(\i*360/\numberOfAngles+\rotationOfMagSphere)+\lengthCoordinateArrow});
  }
  
\foreach \i in {0,...,\numexpr\numberOfAngles-1\relax}{
  \draw[->,ibidark,line width=1pt] ({\r*sin(\i*360/\numberOfAngles+\rotationOfMagSphere)+\shiftX},{\r*cos(\i*360/\numberOfAngles+\rotationOfMagSphere)})--({\r*sin(\i*360/\numberOfAngles+\rotationOfMagSphere)+\shiftX+\lengthCoordinateArrow},{\r*cos(\i*360/\numberOfAngles+\rotationOfMagSphere)});
  }

\draw[->,line width=1pt] (\shiftX,\r+0.5) arc (90:135:2);
\draw[->,line width=1pt] (\shiftX,-\r-0.5) arc (270:315:2);

\def\descriptionHeight{-2.8}

\draw [ar] (2,0) -- (3,0);
\draw [ar] (7,0) -- (8,0);
\node[align=center,ukesec3] at (\shiftX/4,\descriptionHeight/2) {\footnotesize Local Sensor \\ \footnotesize Calibration};
\node[align=center,ukesec3] at (\shiftX*3/4,\descriptionHeight/2) {\footnotesize Gobal \\ \footnotesize Orientation \\ \footnotesize Correction};
\draw [ar] (2,\descriptionHeight) -- (3,\descriptionHeight);
\draw [ar] (7,\descriptionHeight) -- (8,\descriptionHeight);


\node at (0-1.5,2) {a)};
\node at (5-1.5,2) {b)};
\node at (10-1.5,2) {c)};


\node[align=center] at (0,\descriptionHeight) {$\{\bm b^1, \dots, \bm b^N\}$ \\ $\{\bm r^1, \dots, \bm r^N\}$};
\node[align=center] at (5,\descriptionHeight) {$\{\bm B^1, \dots, \bm B^N\}$ \\ $\{\bm r^1, \dots, \bm r^N\}$};
\node[align=center] at (10,\descriptionHeight) {$\{\bm B^1, \dots, \bm B^N\}$ \\ $\{\bm{\hat{r}}_1, \dots, \bm{\hat{r}}_N\}$};
\node[align=center] at (-3.2,\descriptionHeight) {Field Values \\ Sensor Positions};
\node[align=center] at (-3.2,0) {Magnetometer \\ Array};
\end{tikzpicture}

%% file: tikz/5-results_FFPTrajectory.tex
\pgfplotstableread[]{data/FFPtrajectory_t.csv}\Trajectory
\pgfplotstableread[]{data/FFPtrajectoryXYProjection.csv}\TrajectoryXYProjection
\pgfplotstableread[]{data/FFPtrajectoryXZProjection.csv}\TrajectoryXZProjection
\pgfplotstableread[]{data/FFPtrajectoryYZProjection.csv}\TrajectoryYZProjection

\begin{tikzpicture}
\begin{axis}[view={55}{30},
height=5.5cm,
xlabel = {$x$},
ylabel = {$y$},
zlabel = {$z$},
change x base = true, change y base = true, change z base = true,
x unit = m,
x SI prefix = milli, 
y unit = m,
y SI prefix = milli, 
z unit = m,
z SI prefix = milli,
xmin=-0.01,xmax=0.013,ymin=-0.013,ymax=0.013,zmin=-0.03,zmax=0.025,
unit markings=slash space,
colorbar,
colormap/viridis,
colorbar style={
    ylabel= time,
    height=3cm, 
    width = 0.4cm,
    y unit = s,
    change y base = true,
    ytick scale label code/.code={},
    yshift=-0.5cm,
    unit markings=slash space,
    point meta min=0, point meta max=5
    },
grid=major,      
grid style={dashed, gray}, 
]
    \addplot3[scatter,scatter src=explicit] 
    table [x=x,y=y,z=z, meta=t, restrict expr to domain={\coordindex}{0:48} ] {\Trajectory};

    \addplot3[color=gray!50,line width=1.5 pt, opacity=0.5] 
    table [x=x,y=y,z=z,restrict expr to domain={\coordindex}{0:48} ] {\TrajectoryXYProjection};
    \addplot3[color=gray!50,line width=1.5 pt, opacity=0.5] 
    table [x=x,y=y,z=z,restrict expr to domain={\coordindex}{0:48} ] {\TrajectoryXZProjection};
    \addplot3[color=gray!50,line width=1.5 pt, opacity=0.5] 
    table [x=x,y=y,z=z,restrict expr to domain={\coordindex}{0:48} ] {\TrajectoryYZProjection};
\end{axis}
\end{tikzpicture}

%% file: tikz/4-results_fieldPlot.tex
\def\pathFile{data}

\def\h{3cm}
\def\w{3.88*\h}
\def\arrowLength{1.0}
\pgfmathsetmacro\vsep{1.1cm}  
\pgfmathsetmacro\hsep{1.5cm}  

\def\cmin{0.0} 
\def\cmax{0.013} 
\def\discr{21} 
\def\radius{0.045}

\def\cminDiff{0.0} 
\def\cmaxDiff{0.0004} 

\begin{tikzpicture}
\pgfplotstableread[col sep=comma,]{\pathFile/MagSphere_field.csv}\datatableFieldMag 
\pgfplotstableread[col sep=comma,]{\pathFile/MagSphere_field_quiver.csv}\datatableQuiverMag 
\pgfplotstableread[col sep=comma,]{\pathFile/LakeShore.csv}\datatableFieldLake 
\pgfplotstableread[col sep=comma,]{\pathFile/LakeShore_quiver.csv}\datatableQuiverLake 
\pgfplotstableread[col sep=comma,]{\pathFile/diff_field.csv}\datatableFieldDiff 
\pgfplotstableread[col sep=comma,]{\pathFile/diff_field_quiver.csv}\datatableQuiverDiff 

\pgfplotsset{
    sphere/.style = {color=white, dashed, very thick,outer sep=2pt}, 
    }
\begin{groupplot}[
    group style={
       group size=3 by 3,
       vertical sep=\vsep,
       horizontal sep=\hsep,
       group name=fields,
       x descriptions at=edge bottom,
       y descriptions at=edge left,
       },
    scale only axis,
    height=\h,
    width=\h,
    clip mode=individual,
    view={0}{90},
    point meta = explicit,
    mesh/cols=\discr,
        tick align=outside,
        tickpos=left,
        tick style={/pgfplots/major tick length=3pt, thick, black},
        xlabel shift={-6pt},
        ylabel shift={-5pt},
        unit markings=slash space,
        colormap/viridis,
        shader=interp,
        point meta min=\cmin,
        point meta max=\cmax,
    ]


\nextgroupplot[ylabel = {$y$},
               xlabel = {$x$},
               change x base = true, change y base = true,
               x unit = m, x SI prefix = milli,
               y unit = m, y SI prefix = milli,
               ]
    \addplot3[surf,mesh/cols=11,] table [x=PlaneXY_x,y=PlaneXY_y,meta=PlaneXY_f] {\datatableFieldLake};
    \addplot[
        quiver = {
            u = \thisrow{Xyu}, 
            v = \thisrow{Xyv}, 
            scale arrows = \arrowLength,
            update limits=false,
        },
        -stealth,
        ] 
        table [x=PlaneXY_y,y=PlaneXY_x, ] {\datatableQuiverLake};
    \addplot [domain=-180:180, sphere, samples=100] ({\radius*cos(x)},{\radius*sin(x)});

\nextgroupplot[xlabel = {$x$},
               change x base = true, change y base = true,
               x unit = m, x SI prefix = milli,
               ]
    \addplot3[surf] table [x=PlaneXY_x,y=PlaneXY_y,meta=PlaneXY_f] {\datatableFieldMag};
    \addplot[
        quiver = {
            u = \thisrow{Xyu}, 
            v = \thisrow{Xyv}, 
            scale arrows = \arrowLength,
            update limits=false,
        },
        -stealth,
        ] 
        table [x=PlaneXY_y,y=PlaneXY_x, ] {\datatableQuiverMag};
    \addplot [domain=-180:180, sphere, samples=100] ({\radius*cos(x)},{\radius*sin(x)});

\nextgroupplot[
               point meta min=\cminDiff,
               point meta max=\cmaxDiff,
               xlabel = {$x$},
               change x base = true, change y base = true,
               x unit = m, x SI prefix = milli,
               ]
    \addplot3[surf] table [x=PlaneXY_x,y=PlaneXY_y,meta=PlaneXY_f] {\datatableFieldDiff};
    \addplot[
        quiver = {
            u = \thisrow{Xyu}, 
            v = \thisrow{Xyv}, 
            scale arrows = \arrowLength,
            update limits=false,
        },
        -stealth,
        ] 
        table [x=PlaneXY_y,y=PlaneXY_x, ] {\datatableQuiverDiff};
    \addplot [domain=-180:180, sphere, samples=100] ({\radius*cos(x)},{\radius*sin(x)});


\nextgroupplot[ylabel = {$z$},
               xlabel = {$x$},
               change x base = true, change y base = true,
               x unit = m, x SI prefix = milli,
               y unit = m, y SI prefix = milli,
               ]
    \addplot3[surf,mesh/cols=11,] table [x=PlaneXZ_x,y=PlaneXZ_z,meta=PlaneXZ_f] {\datatableFieldLake};
    \addplot[
        quiver = {
            u = \thisrow{PlaneXZ_u},
            v = \thisrow{PlaneXZ_v},
            scale arrows = \arrowLength,
            update limits=false,
        },
        -stealth,
        ] 
        table [x=PlaneXZ_x,y=PlaneXZ_z, ] {\datatableQuiverLake};
    \addplot [domain=-180:180, sphere, samples=100] ({\radius*cos(x)},{\radius*sin(x)});

\nextgroupplot[xlabel = {$x$},
               change x base = true, change y base = true,
               x unit = m, x SI prefix = milli,
               ]
    \addplot3[surf] table [x=PlaneXZ_x,y=PlaneXZ_z,meta=PlaneXZ_f] {\datatableFieldMag};
    \addplot[
        quiver = {
            u = \thisrow{PlaneXZ_u},
            v = \thisrow{PlaneXZ_v},
            scale arrows = \arrowLength,
            update limits=false,
        },
        -stealth,
        ] 
        table [x=PlaneXZ_x,y=PlaneXZ_z, ] {\datatableQuiverMag};
    \addplot [domain=-180:180, sphere, samples=100] ({\radius*cos(x)},{\radius*sin(x)});

\nextgroupplot[
               point meta min=\cminDiff,
               point meta max=\cmaxDiff,
               xlabel = {$x$},
               change x base = true, change y base = true,
               x unit = m, x SI prefix = milli,
               ]
    \addplot3[surf] table [x=PlaneXZ_x,y=PlaneXZ_z,meta=PlaneXZ_f] {\datatableFieldDiff};
    \addplot[
        quiver = {
            u = \thisrow{PlaneXZ_u}, 
            v = \thisrow{PlaneXZ_v}, 
            scale arrows = \arrowLength,
            update limits=false,
        },
        -stealth,
        ] 
        table [x=PlaneXZ_x,y=PlaneXZ_z, ] {\datatableQuiverDiff};
    \addplot [domain=-180:180, sphere, samples=100] ({\radius*cos(x)},{\radius*sin(x)});


\nextgroupplot[ylabel = {$y$},
               xlabel = {$z$},
               change x base = true, change y base = true,
               x unit = m, x SI prefix = milli,
               y unit = m, y SI prefix = milli,
               ]
    \addplot3[surf,mesh/cols=11,] table [x=PlaneYZ_z,y=PlaneYZ_y,meta=PlaneYZ_f] {\datatableFieldLake};
    \addplot[
        quiver = {
            u = \thisrow{quiver_yzv},
            v = \thisrow{quiver_yzu},
            scale arrows = \arrowLength,
            update limits=false,
        },
        -stealth,
        ] 
        table [x=PlaneYZ_z,y=PlaneYZ_y, ] {\datatableQuiverLake};
    \addplot [domain=-180:180, sphere, samples=100] ({\radius*cos(x)},{\radius*sin(x)});

\nextgroupplot[xlabel = {$z$},
               change x base = true, change y base = true,
               x unit = m, x SI prefix = milli,
               ]
    \addplot3[surf] table [x=PlaneYZ_z,y=PlaneYZ_y,meta=PlaneYZ_f] {\datatableFieldMag};
    \addplot[
        quiver = {
            u = \thisrow{quiver_yzv},
            v = \thisrow{quiver_yzu},
            scale arrows = \arrowLength,
            update limits=false,
        },
        -stealth,
        ] 
        table [x=PlaneYZ_z,y=PlaneYZ_y, ] {\datatableQuiverMag};
    \addplot [domain=-180:180, sphere, samples=100] ({\radius*cos(x)},{\radius*sin(x)});

\nextgroupplot[
               point meta min=\cminDiff,
               point meta max=\cmaxDiff,
               xlabel = {$z$},
               change x base = true, change y base = true,
               x unit = m, x SI prefix = milli,
               ]
    \addplot3[surf] table [x=PlaneYZ_z,y=PlaneYZ_y,meta=PlaneYZ_f]  {\datatableFieldDiff};
    \addplot[
            quiver = {
            u = \thisrow{quiver_yzv},
            v = \thisrow{quiver_yzu},
            scale arrows = \arrowLength,
            update limits=false,
        },
        -stealth,
        ] 
        table [x=PlaneYZ_z,y=PlaneYZ_y, ] {\datatableQuiverDiff};
    \addplot [domain=-180:180, sphere, samples=100] ({\radius*cos(x)},{\radius*sin(x)});

\end{groupplot}

\tikzset{
        capt/.style = {text centered, text width=\h, xshift=0.0cm, anchor=south, 
                        draw = ibidark, line width = 3pt, rounded corners=5pt, 
                        fill = ibilight!25,
                        minimum height=32pt, inner sep=0pt, font=\Large},
    }

\node[above = 10pt of fields c1r1] {Single Hall Sensor};
\node[above = 10pt of fields c2r1] {Magnetometer Array};
\node[above = 10pt of fields c3r1] {Difference\vphantom{g}};

\path (fields c1r1) --node[midway] {\Huge $\bm -$} (fields c2r1);
\path (fields c1r2) --node[midway] {\Huge $\bm -$} (fields c2r2);
\path (fields c1r3) --node[midway] {\Huge $\bm -$} (fields c2r3);
\path (fields c2r1) --node[midway] {\Huge $\bm =$} (fields c3r1);
\path (fields c2r2) --node[midway] {\Huge $\bm =$} (fields c3r2);
\path (fields c2r3) --node[midway] {\Huge $\bm =$} (fields c3r3);

\path (fields c1r3.south) -- node (cbPos) {} (fields c2r3.south);
\node[below = of cbPos,xshift=-0.7mm,yshift=-0.3cm] (cb) {
\pgfplotscolorbardrawstandalone[
    colormap/viridis, 
    colorbar horizontal,
    point meta min=\cmin,
    point meta max=\cmax,
    scale only axis,
    colorbar style={
        colorbar shift/.style={xshift=0.075*\h},
        height=0.5cm,
        width=2*\h+\hsep,
        xticklabel style = {xshift=0.0cm,yshift=0.0cm},
        x tick style= {color=black, thick},
        extra x tick style={tickwidth=0pt},
        change x base = true,
        xlabel = {$\lVert \bm B \rVert_2$},
        x unit = T, x SI prefix = milli,
        unit markings=slash space,
        xtick={0, 0.002, 0.004, 0.006, 0.008, 0.010, 0.012, 0.014},
    },
    ]};
\node[at = (cb -| fields c3r3),xshift=0mm,yshift=0cm] (cbDiff) {
\pgfplotscolorbardrawstandalone[
    colormap/viridis, 
    colorbar horizontal,
    point meta min=\cminDiff,
    point meta max=\cmaxDiff,
    colorbar style={
        colorbar shift/.style={xshift=0.075*\h},
        height=0.5cm,
        width=1.0*\h,
        xticklabel style = {xshift=0.0cm,yshift=0.0cm},
        x tick style= {color=black, thick},
        extra x tick style={tickwidth=0pt},
        change x base = true,
        xlabel = {$\lVert \bm B \rVert_2$},
        x unit = T, x SI prefix = milli,
        unit markings=slash space,
        xtick={0, 0.0001, 0.0002, 0.0003,0.0004},
    },
    ]};
\end{tikzpicture}

%% file: tikz/3-results_fieldAccuracy_upright.tex
\pgfplotstableread[]{data/OnLine_LakeShore.txt}\LakeShore
\pgfplotstableread[]{data/OnLine_MagSphere.txt}\MagSphere %
\pgfplotstableread[]{data/OnLine_error.txt}\Error 

\pgfplotsset{
  layers/axis lines on top/.define layer set={
    axis background,
    axis grid,
    axis ticks,
    axis tick labels,
    pre main,
    main,
    axis lines,
    axis descriptions,
    axis foreground,
  }{/pgfplots/layers/standard},
}

\def\w{3.2cm}
\def\h{2cm}
\pgfmathsetmacro\vsep{1.3cm}  
\pgfmathsetmacro\hsep{1.5cm}

\def\fontSize{\scriptsize}
  
\begin{tikzpicture}[node distance = -0.1cm and -0.05cm]

\begin{groupplot}[
    group style={
       group size=2 by 3,
       vertical sep=\vsep,
       horizontal sep=\hsep,
       group name=gPlot,
       ylabels at=edge left,
       },
            set layers=axis lines on top, 
            axis background/.style={fill=white},
            change x base = true,
            change y base = true,
            x unit = m,
            x SI prefix = milli, 
            y unit = T, 
            compat=1.3,
            y label style={yshift=-0.15cm}, 
            label style = {font=\fontSize},
            tick label style={font=\fontSize},
            xtick={-0.04,-0.02,0,0.02,0.04},
            enlargelimits=false,
            tick align = outside,
            tickpos=left,
            grid=major,
            tick style={major grid style={line width=0.4pt,dashed,gray},
                },
            legend style={font=\scriptsize, line width=1pt,
                fill=white, draw=none,
                /tikz/every even column/.append style={column sep=0.8cm},inner sep=2mm,rounded corners=2pt,
                   minimum height=4pt,
                /tikz/every even column/.append style={column sep=0.6cm},
                legend image post style={xscale=0.6},
                   },
            legend columns = 3,
            legend cell align={left},
            height=\h,
            width=\w,
            scale only axis,
    ]

\nextgroupplot[
    y SI prefix = milli,
    x unit = m,
    x SI prefix = milli,
    xlabel={$x$},
    ylabel={$\lVert \bm B \rVert_2$},
    legend style={at={(1.15,1.1)}, anchor=south, legend columns=3},
    ]

    \addplot[ibidark,smooth,thick,line width=1.5pt] table [x=Range,y=Bx,col sep=comma]{\LakeShore};
    \addlegendentry{Single Hall Sensor};

    \addplot[ukesec2,smooth,thick,line width=1.5pt,dashed] table [x=Range,y=Bx,col sep=comma]{\MagSphere};
    \addlegendentry{Magnetometer Array};

    \addlegendimage{mark=none, ukesec4, thick, line width=1.0pt};
    \addlegendentry{Error};

\nextgroupplot[
    y SI prefix = micro,
    y unit = T,
    x unit = m,
    x SI prefix = milli,
    xlabel={$x$},
    ylabel={$\lVert \bm B_\text{diff} \rVert_2$},
    ymin = 0,
    ymax = 0.0003,
    ]

    \addplot[ukesec4,smooth,ultra thick,line width=1.5pt,] table [x=Range,y=eX]{\Error};

\nextgroupplot[
    xlabel={$y$},
    ylabel={$\lVert \bm B \rVert_2$},
    y SI prefix = milli,
    y unit = T,
    x unit = m,
    x SI prefix = milli,
    ]

    \addplot[ibidark,smooth,thick,line width=1.5pt] table [x=Range,y=By,col sep=comma]{\LakeShore};

    \addplot[ukesec2,smooth,thick,line width=1.5pt,dashed] table [x=Range,y=By,col sep=comma]{\MagSphere};

\nextgroupplot[
    y SI prefix = micro,
    y unit = T,
    x unit = m,
    x SI prefix = milli,
    xlabel={$y$},
    ylabel={$\lVert \bm B_\text{diff} \rVert_2$},
    ymin = 0,
    ymax = 0.0003,
    ]

    \addplot[ukesec4,smooth,ultra thick,line width=1.5pt,] table [x=Range,y=eY]{\Error};
    
\nextgroupplot[
    y SI prefix = milli,
    y unit = T,
    x unit = m,
    x SI prefix = milli,
    xlabel={$z$},
    ylabel={$\lVert \bm B \rVert_2$},
    ]
    
    \addplot[ibidark,smooth,thick,line width=1.5pt] table [x=Range,y=Bz,col sep=comma]{\LakeShore};

    \addplot[ukesec2,smooth,thick,line width=1.5pt,dashed] table [x=Range,y=Bz,col sep=comma]{\MagSphere};
    
\nextgroupplot[
    y SI prefix = micro,
    y unit = T,
    x unit = m,
    x SI prefix = milli,
    xlabel={$z$},
    legend style={at={(1,1)},anchor=south east},
    ylabel={$\lVert \bm B_\text{diff} \rVert_2$},
    ymin = 0,
    ymax = 0.0003,
    ]

    \addplot[ukesec4,smooth,ultra thick,line width=1.5pt,] table [x=Range,y=eZ]{\Error};
\end{groupplot}

\end{tikzpicture}